\documentclass[sigplan, dvipsnames, nonacm, pdftex]{acmart}
\usepackage{xcolor}
\usepackage{algorithmic}
\usepackage{graphicx}
\usepackage{textcomp}
\usepackage{url}
\usepackage{fancyhdr}
\usepackage{hyperref}
\usepackage[utf8]{inputenx}
\usepackage[T1]{fontenc}
\usepackage{amsthm, amsmath, amsfonts}
\usepackage{algorithmic}
\usepackage{graphicx}
\usepackage{textcomp}
\usepackage{color, colortbl} 
\usepackage{multirow}
\usepackage{enumitem}
\usepackage{multicol}
\usepackage[altpo]{backnaur}
\usepackage{booktabs} 
\usepackage{listings}
\usepackage{tikz}
\usepackage{float}
\usepackage{url}
\usepackage{hyperref}
\usepackage{mathpartir}
\usepackage{fancyhdr}
\usepackage{pifont}
\usepackage{xspace}
\usepackage[capitalize]{cleveref}
\usepackage{caption}
\usepackage{array}
\usepackage{boldline}
\usepackage{soul}
\usepackage{listings}
\newif\ifhighlightrev
\highlightrevfalse
\newcommand{\rev}[1]{\ifhighlightrev{\color{black}#1}\else{#1}\fi}
\newcommand{\del}[1]{}

\definecolor{bug}{HTML}{FF6666}
\definecolor{proof}{HTML}{006600}
\newcolumntype{L}[1]{>{\raggedright\let\newline\\\arraybackslash\hspace{0pt}}m{#1}}
\newcolumntype{C}[1]{>{\centering\let\newline\\\arraybackslash\hspace{0pt}}m{#1}}
\newcolumntype{R}[1]{>{\raggedleft\let\newline\\\arraybackslash\hspace{0pt}}m{#1}}

\usepackage{fontawesome5}
\newcommand{\bug}{{\scriptsize{\faBug}}}
\newcommand{\secure}{{\scriptsize{\faGrin[regular]}}}
\newcommand{\timeout}{{\scriptsize{\faClock}}}
\newcommand{\falsealert}{{\scriptsize{\faExclamationTriangle}}}
\newcommand{\graycell}{{\cellcolor{lightgray}}}

\usepackage{tablefootnote}

\makeatletter
\let\old@lstKV@SwitchCases\lstKV@SwitchCases
\def\lstKV@SwitchCases#1#2#3{}
\makeatother
\usepackage{lstlinebgrd}
\makeatletter
\let\lstKV@SwitchCases\old@lstKV@SwitchCases

\newif\ifdraft
\drafttrue
\newcommand{\qt}[1]
{\ifdraft{\textsf{\footnotesize\color{magenta}[#1 --Qinhan]}}\fi}

\newcommand{\mengjia}[1]
{\ifdraft{\textsf{\footnotesize\color{brown}[Mengjia: #1]}}\fi}
\newcommand{\mengjiatext}[1]
{\ifdraft{\color{brown}#1}\else#1\fi}

\newcommand{\thomas}[1]
{\ifdraft{\textsf{\footnotesize\color{red}[Thomas: #1]}}\fi}

\newcommand{\cameraready}[1]{\ifdraft{\color{black}#1}\else{#1}\fi}

\newcommand{\name}{Contract Shadow Logic\xspace}
\newcommand{\swcheck}{contract constraint check\xspace}
\newcommand{\hwcheck}{leakage assertion check\xspace}

\renewcommand{\paragraph}[1]{\noindent\textbf{#1 \xspace}}

\begin{document}

\title{RTL Verification for Secure Speculation Using \name}


\author{Qinhan Tan}
\authornote{Two co-first authors contributed equally to this research.}
\email{qinhant@princeton.edu}
\orcid{0000-0003-2475-3675}
\affiliation{%
  \institution{Princeton University}
  \city{Princeton}
  \state{New Jersey}
  \country{USA}
}

\author{Yuheng Yang}
\authornotemark[1]
\email{yuhengy@mit.edu}
\orcid{0000-0001-8695-5139}
\affiliation{%
  \institution{MIT CSAIL}
  \city{Cambridge}
  \state{Massachusett}
  \country{USA}
}

\author{Thomas Bourgeat}
\email{thomas.bourgeat@epfl.ch}
\orcid{0000-0002-8468-8409}
\affiliation{%
  \institution{EPFL}
  \city{Lausanne}
  \country{Switzerland}
}

\author{Sharad Malik}
\email{sharad@princeton.edu}
\orcid{0000-0002-0837-5443}
\affiliation{%
  \institution{Princeton University}
  \city{Princeton}
  \state{New Jersey}
  \country{USA}
}

\author{Mengjia Yan}
\email{mengjiay@mit.edu}
\orcid{0000-0002-6206-9674}
\affiliation{%
  \institution{MIT CSAIL}
  \city{Cambridge}
  \state{Massachusett}
  \country{USA}
}






\begin{abstract}
Modern out-of-order processors face speculative execution attacks.
Despite various proposed software and hardware mitigations to prevent such attacks, new attacks keep arising from unknown vulnerabilities.
Thus, a formal and rigorous evaluation of the ability of hardware designs to deal with speculative execution attacks is urgently desired.

This paper proposes a formal verification technique called \emph{\name} that can considerably improve RTL verification scalability while being applicable to different defense mechanisms.
In this technique, we leverage {computer architecture} design insights to 
improve verification performance for checking security properties formulated as software-hardware contracts for secure speculation.
 Our verification scheme is accessible to computer architects and requires minimal formal-method expertise.

We evaluate our technique on multiple RTL designs, including three out-of-order processors.
The experimental results demonstrate that our technique exhibits a significant advantage in finding attacks on insecure designs and deriving complete proofs on secure designs, when compared to the baseline and two state-of-the-art verification schemes, LEAVE and UPEC.

\end{abstract}
\maketitle


\pagestyle{plain}

\lst@Key{numbers}{none}{%
    \def\lst@PlaceNumber{\lst@linebgrd}%
    \lstKV@SwitchCases{#1}%
    {none:\\%
     left:\def\lst@PlaceNumber{\llap{\normalfont
                \lst@numberstyle{\thelstnumber}\kern\lst@numbersep}\lst@linebgrd}\\%
     right:\def\lst@PlaceNumber{\rlap{\normalfont
                \kern\linewidth \kern\lst@numbersep
                \lst@numberstyle{\thelstnumber}}\lst@linebgrd}%
    }{\PackageError{Listings}{Numbers #1 unknown}\@ehc}}
\makeatother

\lstdefinestyle{mystyle}{
  basicstyle=\ttfamily\small,
  keywordstyle={[1]\color{Magenta}},
  keywordstyle={[2]\color{BrickRed}},
  keywordstyle={[3]\color{RoyalBlue}},
  commentstyle=\color{OliveGreen},
  stringstyle=\color{Orange},
  numbers=left,
  numberstyle=\tiny\color{gray},
  numbersep=3pt,
}
\lstdefinelanguage{mylang} {
    morekeywords={[1]input, output, always, posedge, begin, end, if, else, assume, assert, property, case, then, endif, initial},
    morekeywords={[2]},
    morekeywords={[3]},
    morekeywords={[4]},
    sensitive=false,
    morecomment=[l]{//},
}
\lstset{style=mystyle}


\section{Introduction}
\label{sec:intro}

{Speculative execution} in modern out-of-order (OoO) processors has given rise to various speculative execution {attacks}, including Spectre~\cite{Kocher2018spectre}, Meltdown~\cite{Lipp2018meltdown}, L1TF~\cite{website:l1tf_intel}, and LVI~\cite{van2020lvi}. 
These attacks exploit the microarchitectural side effects of transient instructions, i.e., the instructions that are executed speculatively due to misspeculation and squashed later.
Prior work has demonstrated that speculative execution attacks can leak secrets via a wide range of microarchitectural structures, including caches~\cite{Kocher2018spectre, kiriansky2018speculative, koruyeh2018spectre}, TLBs~\cite{gras2018translation}, branch predictors~\cite{chowdhuryy2021leaking} and functional units~\cite{bhattacharyya2019smotherspectre}.

Even worse, as the microarchitecture becomes ever more complex, anticipating
speculative execution {vulnerabilities} seems to pose a Sisyphean challenge. 
For example, Foreshadow~\cite{van2018foreshadow} identified the L1TF vulnerability, showcasing cross-domain leakage of data residing in the L1 cache.
Though the L1TF vulnerability was quickly patched by Intel by flushing L1 data cache upon context switch~\cite{website:l1tf_intel}, such mitigation was insufficient to block small variations of the attack, due to the existence of various other structures inside the L1, including the line fill buffer, the store buffer, the load port, etc.
These structures were exploited to leak information by a batch of follow-on attacks called Microarchitecture Data Sampling Attacks (MDS)~\cite{schwarz2019zombieload, van2019ridl}.
The cat-and-mouse game described above calls for rigorous approaches to ensure 
the security of processors.


To this end, this paper aims to develop a formal verification scheme for secure speculation
on out-of-order processors working at the \textit{Register Transfer Level (RTL)}.
\cameraready{
As hardware design is a lengthy procedure and involves multiple steps, including early-stage high-level modeling and later-stage Register Transfer Level (RTL) implementation, we need verification tools at each stage to offer security guarantees.
Several prior work, such as Pensieve~\cite{yang2023pensieve}, CheckMate~\cite{trippel2018checkmate}, and in Guarnieri et al.~\cite{guarnieri2021hardware}, focus on higher-level abstract models to assist security verification at early design stage.
However, higher-level models are prone to overlooking small but crucial processor structures, like the buffers exploited by the MDS attacks. 
Thus, we aim to design verification tools that can operate at the RTL level at a later design stage when all microarchitectural details are in place.
We envision our tool can work complementarily with existing early-stage verification tools for the whole hardware design flow.
}


\subsection{Blueprint for Systematic and Scalable RTL Security Verification}

Existing RTL formal verification schemes for secure speculation pose scalability challenges.
Most techniques only work effectively on simple designs such as in-order processors, but struggle to cope with architecturally complex designs like out-of-order processors, leading to timeout.
Making a verification scheme work on an out-of-order processor requires the verification engineer to spend significant manual effort to customize security specifications of subcomponents and try to find invariants that help the unbounded proof to go through. 
This process is very challenging, especially as it requires the engineers to have security expertise and in-depth knowledge of formal methods, computer architecture, and the specifics of the design studied.
Moreover, the verification harness that works on one design cannot be easily transferred to another design without requiring new creative insights. 

For example, the state-of-the-art verification work UPEC~\cite{fadiheh2020formal} successfully proves a secure speculation scheme on BOOM~\cite{zhaosonicboom}, highlighting a breakthrough along this research direction.\footnote{There are multiple versions of UPEC dealing with different security issues~\cite{schmitz2023upec, deutschmann2023scalable}.
In this paper, unless otherwise stated, we refer to \cite{fadiheh2020formal} which tackles speculative vulnerabilities in RTL implementations.}
However, a careful examination of their scheme shows that the verification methodology
is depending on the specific conservative defense mechanism being verified (which is to prevent speculative load instructions to execute before all preceding branch instructions in the ROB are resolved).
To verify other defenses such as DoM~\cite{sakalis2019efficient} and STT~\cite{yu2019speculative}, a UPEC user will face extensive challenges and need to invest additional effort to extend the set of invariants. 

Alternatively, recent work (LEAVE~\cite{wang2023specification}) proposed to reduce the amount of manual effort needed for RTL verification of secure speculation schemes using automatic invariant search techniques.
Those techniques demonstrated great effectiveness on certain in-order pipelined processors, however we found that the approach tends to have a hard time on simple out-of-order processors (Section~\ref{sec:eval:compare}).

To summarize, the primary research challenge here is to improve verification scalability to more evolved architectures and reduce the amount of manual and non-reusable efforts required to complete the RTL security verification.

In this paper, we try to make progress towards reusable and more scalable verification schemes by leveraging computer architecture insights.
Our verification approach gets the computer architects who design the processor involved in the formal verification loop: we are asking them to design the shadow logic machinery instrumental to our verification task, from which we can call the verification tools to formally check the security property.
Compared to writing logical invariants, writing this shadow logic machinery is a task much closer to the standard activity of computer architects.

\subsection{This Paper: Contract Shadow Logic}

The security property to be checked in secure speculation is expressed as a software-hardware contract~\cite{guarnieri2021hardware, cauligi2022sok},
which states, ``if a software program satisfies a constraint, then the hardware ensures executing the program will not leak secrets.''
Checking this statement involves two operations.
We refer to the operation that checks the software against a constraint as the \textbf{\swcheck}, and refer to the operation that checks the hardware for detecting {information leakage} as the \textbf{\hwcheck}.
Each of the two checks is performed on a pair of state machines (or machines for short). 
The \swcheck compares the executions of two simple machines which implement the ISA and execute each instruction in one-cycle (henceforth referred to as {single-cycle machines} or {ISA machines}).
Similarly the \hwcheck compares the executions of two copies of the out-of-order processor to check for secret leakage.
The baseline verification scheme examines the \textit{four} state machines cycle by cycle.

We propose a verification scheme, called {\name}.
The key insight is that we can \textit{perform both the \swcheck and the \hwcheck on a single pair of out-of-order machines}, and thus eliminate the pair of single-cycle machines in the verification scheme.
Specifically, assuming the out-of-order processor is correctly implementing the ISA semantics, the single-cycle machine's execution trace can be recovered from the out-of-order processor's \textit{committed} instruction sequences.
We implement this ISA trace extraction mechanism as shadow logic (which is added to assist verification, does not interfere with the original design, and can be safely removed before synthesis).


We have implemented our verification scheme using the commercial RTL verification tool, JasperGold~\cite{website:jaspergold} and use it to verify one in-order processor, three out-of-order processors, and five defense augmentations.
We provide experimental results to compare our scheme with the baseline and two state-of-the-art verification schemes, LEAVE~\cite{wang2023specification} and UPEC~\cite{fadiheh2020formal}.
\cameraready{Through our results, we demonstrate the following:}
First, across all secure designs, the baseline scheme cannot provide any security proofs within 7 days, while, our scheme can prove the security for all of them.
Second, while LEAVE cannot automatically find bugs or prove security for our in-house simplified out-of-order processor, our new scheme is able to do so.
Third, we show the ability of our method to find attacks arising from different mis-speculation sources on the BOOM~\cite{zhaosonicboom} processor, while UPEC can only detect attacks caused by pre-identified mis-speculation sources.
\cameraready{However, we would like to point out that scalability continues to be a challenge in formally verifying secure speculation of hardware designs, and one will need additional architectural insights to assist formal verification to scale to more complex out-of-order processors.}


In summary, this paper makes the following contributions:
\begin{enumerate}[leftmargin=*]
    \item We propose a verification scheme, called \name.
    Our scheme leverages computer architecture insights to improve the scalability of RTL security verification for secure speculation while maintaining broad applicability (\cref{sec:insight}).
    \item We identify two important requirements to construct correct shadow logic for checking software-hardware contracts (\cref{sec:main}).
    \item We demonstrate the efficacy of our scheme on four processors and conduct a detailed comparison to showcase the advantages of our scheme compared to the baseline and two state-of-the-art verification schemes, i.e., LEAVE and UPEC (\cref{sec:eval}).
\end{enumerate}

\section{Background}
\label{sec:background}


\subsection{Speculative Execution Attacks and Defenses}
\label{sec:background:spec}

Speculative execution attacks exploit the microarchitectural side effects of transient instructions.
Transient instructions can be introduced via a wide range of speculation features, including branch history prediction~\cite{kiriansky2018speculative}, branch target prediction~\cite{Kocher2018spectre}, return address prediction~\cite{koruyeh2018spectre,maisuradze2018ret2spec}, store-to-load forwarding~\cite{horn2018speculative},
value prediction~\cite{vicarte2021opening}, and 4K aliasing~\cite{canella2019fallout}.
The side effects of transient instructions are diverse, including modifying the cache states~\cite{Kocher2018spectre, Lipp2018meltdown,koruyeh2018spectre,maisuradze2018ret2spec,kiriansky2018speculative}, TLBs~\cite{gras2018translation}, Line Fill Buffer~\cite{schwarz2019zombieload}, and DRAMs~\cite{tobah2022spechammer}.



Speculative execution vulnerabilities can be mitigated via software and hardware methods.
In software, existing mitigations typically involve placing \texttt{fence} instructions after sensitive branch instructions to delay the execution of the subsequent instructions until the branch is resolved~\cite{amd2023software,intel2021intel}.
Other software approaches involve replacing branch instructions with \texttt{cmov} instructions and using bit masking of the index used for array lookups~\cite{carruth2018speculative}.
Hardware mitigation mechanisms focus on restricting speculative execution by either selectively delaying the execution of instructions that can potentially transmit secrets via side channels~\cite{sakalis2019efficient,yu2019speculative, loughlin2021dolma, weisse2019nda} 
or making it difficult for attackers to observe the microarchitectural side effects of speculative instructions~\cite{yan2018invisispec,ainsworth2021ghostminion, ainsworth2020muontrap, khasawneh2019safespec}.



\subsection{SW-HW Contracts for Secure Speculation}
\label{sec:background:contract}


As speculative execution vulnerabilities can be mitigated via either software or hardware methods, it becomes challenging to cleanly and explicitly reason about security properties a defense should achieve.
The community has been using \textbf{software-hardware contracts} to define security properties precisely related to speculative execution vulnerabilities.

Guarnieri et al.~\cite{guarnieri2021hardware} proposed the first set of formalized contracts by defining which program executions a side channel adversary can distinguish.
Specifically, a contract states that: ``if a program satisfies a specified \textbf{software constraint}, then the hardware needs to achieve a \textbf{noninterference} property to ensure no secrets are being leaked.''
Formally, a contract can be written as follows:

\begin{equation}    
\begin{aligned}
&\forall P, M_{pub}, M_{sec}, M_{sec}', \\
\text{\textbf{if}}\quad & O_{ISA}(P,M_{pub},M_{sec})=O_{ISA}(P,M_{pub},M_{sec}') \\
\text{\textbf{then}}\quad & O_{\mu Arch}(P, M_{pub},M_{sec}) = O_{\mu Arch}(P, M_{pub},M_{sec}') & 
\end{aligned}
\label{eq:4trace}
\end{equation}

The formula states that when running a program $P$ with public memory $M_{pub}$ and secret memory $M_{sec}$ on a microarchitecture processor $\mu Arch$, if the program $P$ satisfies an indistinguishability software observation denoted by $O_{ISA}$, then the actual observation when executed on the hardware denoted as $O_{\mu Arch}$ remains indistinguishable.

Note that \cref{eq:4trace} does not describe a single contract but a family of contracts.
The equation is parameterized by different execution modes at the software level and different observation functions.
The scheme proposed in this paper supports the \textit{sequential} execution mode where the software constraint considers traces when executing the program sequentially (i.e., not speculatively).
For example, under a so-called sandboxing contract, the software constraint is that executing the program sequentially will not have secrets loaded into the register file.
The $O_{ISA}$ of sandboxing includes the data written to registers of every committed load.
In contrast, in the constant-time programming contract, the software constraint is that executing the program sequentially will not use secrets as addresses of instruction or data memory accesses or as operands to hardware modules whose timing depends on the input.
This matches the constant-time programming discipline~\cite{guarnieri2021hardware}.
The $O_{ISA}$ of this contract includes the branch condition, memory address and multiplication operands of committed instructions.

The microarchitecture observation ($O_{\mu Arch}$) considered in this paper includes the address sequence
at the memory bus and the commit time of every committed instruction, which is commonly used in prior works~\cite{oleksenko2022revizor, yang2023pensieve} to reflect cache and timing side channels.

\subsection{RTL Verification using Model Checking}


Model checking is widely used in verifying hardware designs, including processors~\cite{berezin2002verification} and accelerators~\cite{huang2018instruction}.
It is available as both open-source tools (e.g., AVR~\cite{goel2020avr}) and commercial tools (e.g., Cadence Jaspergold~\cite{website:jaspergold}) that directly work on RTL designs.
These tools are highly automated, where the user only needs to provide the hardware source code instrumented with the property to be checked.
Existing tools have flexible support to express various properties via instrumenting hardware code using a design-friendly language, e.g., System Verilog Assertions (SVA). SVA allows specifying (i)  assumptions (using the \texttt{assume} keyword) that describe system assumptions such as constraints on the inputs and (ii) assertions (using the \texttt{assert} keyword) to specify the property that the model checker is required to check, similar to assertions in C/Python. 

As model checking effectively considers all possible input sequences which can be of unbounded length, it provides an {unbounded-length proof} or a complete proof. 
When the property does not hold at some reachable state, model checking reports an input-sequence trace, a {counterexample}, for which the property fails.
{Bounded model checking} (BMC) can be used to perform verification for {all possible input sequences of length bounded by some user-provided parameter $k$} - this effectively checks the property for all states reachable within $k$ steps from the initial state \rev{which is obviously not as strong as an unbounded-length proof}.

The information leakage security property is a hyperproperty, i.e., it needs reasoning over two traces (corresponding to two different values of the secret) to prove or disprove it. This can be handled by constructing a model checking instance with two copies of the design (with different values of the secret), and constructing a property comparing traces over the two copies.

\subsection{Shadow Logic}
\label{sec:background:shadow}


When verifying complex RTL code, it is often necessary to introduce auxiliary code to assist verification by obtaining or deriving extra information from the current state of the system.
Such auxiliary code is called {shadow logic} (or ghost code in software verification~\cite{leino2010dafny}). 
This logic \textit{monitors} the design and assumes and asserts that certain conditions hold.

For example, consider a verification task that checks whether a processor correctly implements its ISA specification.
For every instruction that updates the register file, we need to check that when the instruction commits, the target register holds the value specified by the ISA. 
However, due to complex renaming logic in out-of-order processors, it is a nontrivial task to determine which register in the physical register file is the target register.
Prior work~\cite{reid2016end} customizes shadow logic according to the renaming logic to correctly locate the target register as part of this checking.

It is important that this shadow logic only \emph{monitors} the design and does \textit{not} make any modifications that change the behavior of the system. 
\cameraready{
It is common practice to implement shadow logic for verification purposes under some macro.
This enables this logic to be stripped out during design compilation and thus, shadow logic is not present in the final production RTL. 
}

\section{Threat Model}


Recall from \cref{sec:background:contract}, security properties for secure speculation can be formulated as software-hardware contracts.
Such contracts explicitly state the responsibilities of the software and the hardware sides to achieve an overall non-interference property~\cite{clarkson2010hyperproperties}.
In this paper, we consider the following threat model.

\begin{enumerate}[leftmargin=*]
    \item We assume a system whose data memory is split into two parts: \emph{public} and \emph{secret}. 
    An attacker cannot directly access the secret region.
    The attacker's goal is to infer any secret value through microarchitectural side-channel attacks.
    Non-digital side-channel attacks, such as power side channels~\cite{hertzbleed, poweranalysisattacks}, are out of the scope.
    
    \item The system runs a program that satisfies certain software constraints.
    These constraints prevent the secret from being leaked via ISA state.
    For example, at the ISA level, the program cannot load secret data into registers (in the sandboxing contract), or it needs to follow the constant-time programming discipline.

    \item The attacker is able to gain cycle-level observation of certain micro-architectural variables in the system, including the address sequence transmitted on the memory bus and the commit time of instructions.
    
\end{enumerate}

\section{Motivation and Insights}
\label{sec:insight}

In this section, we describe the insights that motivate the design of our verification scheme.
We start by presenting a baseline verification scheme for secure speculation and we analyze the performance bottlenecks encountered.
We then point out the optimization insight that is architecture-friendly and leads us to the design of \name.


\subsection{The Baseline Verification Scheme}
\label{sec:insight:baseline}

Verifying the software-hardware contract described in Section~\ref{sec:background:contract} requires performing \textit{two} kinds of equivalence checks that involve \textit{four} traces.
The equivalence check at line 2 in \cref{eq:4trace} is the \textit{\swcheck} to determine whether a program $P$ needs to be protected (we say that $P$ is valid).
The equivalence check at line 3 is the \textit{\hwcheck} to determine whether the microarchitecture processor $\mu Arch$ has a correct defense mechanism for the given program.

\begin{figure}[t]
    \centering
    \includegraphics[width=\columnwidth]{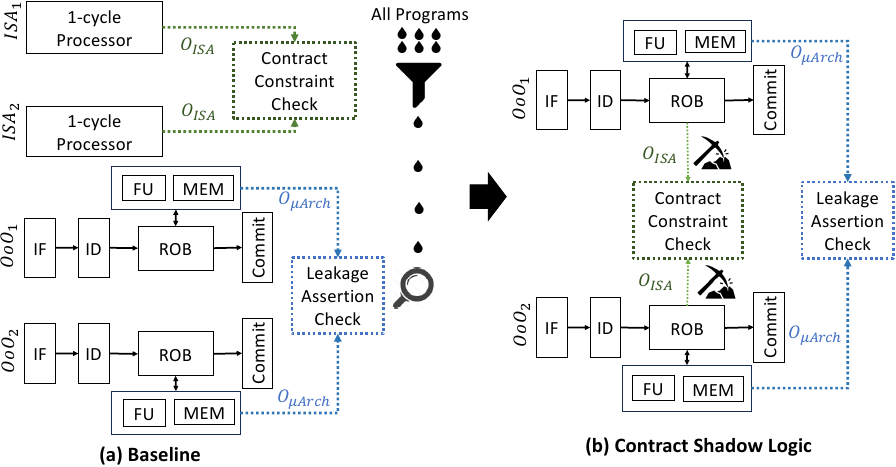}
    \caption{Overview of \name.}
    \label{fig:overview}
\end{figure}


The security property (\cref{eq:4trace}) can be tentatively verified using the scheme visualized in \cref{fig:overview}(a).
First, two copies of a single-cycle machine are instantiated as two separate RTL modules.
The two modules have identical RTL code with the instruction memory and public data memory initialized with the same content ($P$ and $M_{pub}$) and different secret memory ($M_{sec}$), i.e., two secret memories differ in at least one location.
The two instances of single-cycle machines execute instructions sequentially, finishing one instruction per cycle.
These two copies execute in a lock-step manner, allowing the verification scheme to derive the $O_{ISA}$ trace at each cycle and compare them for equivalence to enforce the \swcheck.
\cameraready{The comparison results serve as a constraint to filter out invalid programs for the out-of-order processor being verified.
The model checker will discard the programs that generate the traces violating the constraint.}

Similarly, the RTL for the out-of-order processor is duplicated to create two instances, which only differ in the secret memory.
The program memory $P$ and the initial value of the public data memory $M_{pub}$ are the same as in the single-cycle machine. 
The verification scheme compares the microarchitecture execution traces $O_{\mu Arch}$ from the two machines at each cycle for the \hwcheck.


This scheme runs the \swcheck and \hwcheck in parallel at each cycle, allowing for conducting both bounded and unbounded checking.

\subsection{Architecture-driven Insights for Verification}

As discussed in \cref{sec:intro}, the community faces serious challenges in coming up with verification schemes that are both reusable for verifying different defense designs and, meanwhile, scalable enough to be able to verify out-of-order processors.
In this paper, we aim to tackle the research challenge of designing a reusable verification scheme while achieving reasonable scalability, where the verification scheme can handle some out-of-order processors without requiring to come up with custom inductive invariants. 
A baseline verification scheme employs four state machines (for the four processors) which entails a non-trivial amount of redundancy among them, we aim to explore the possibility of reducing this redundancy.

As mentioned in Section~\ref{sec:intro}, if an out-of-order processor functionally correctly implements its ISA specification, the ISA trace can be reconstructed from the out-of-order processor's committed instruction sequence.
In other words, the single-cycle machine is contained in the out-of-order machine, including the complex instruction decoding, the execution logic, as well as registers/memory state.
This insight leads to our verification scheme, introducing shadow logic to implement the \swcheck directly on the out-of-order processors, as shown in \cref{fig:overview}(b).
With this, we can reduce a four-machine verification problem (the baseline) to a two-machine problem with some additional shadow logic -- this reduction in logic could potentially lead to better scalability.


\section{\name}
\label{sec:main}


{
In this section, we present the main components of \name. 
We first describe the basic component of our shadow logic that extracts the ISA traces from out-of-order processors.
We then discuss the challenges in making the shadow logic correctly verify software-hardware contracts, and finally provide our solution adressing these challenges.
}

\subsection{Shadow Logic to Extract ISA Traces}

An important task of the shadow logic is to monitor the commit stage of the out-of-order processor and extract information for every committed instruction to derive the ISA trace.
The shadow logic needs to be implemented according to the observation function ($O_{ISA}$).
For example, if checking for the constant-time contract, the \swcheck needs to compare the addresses of committed memory access instructions and the condition values of committed branch instructions.
Different processor implementations may store this information in different places.
In the best case, the information is already present in the ROB and can be directly obtained.


In more complex cases, the information may be spread across multiple modules, such as inside the register files and load/store queues, and in some cases the information might not even be readily available at commit-time anymore.
{
This is the place where we can leverage and involve the architect's knowledge and expertise.
}
Specifically, to handle these cases, the user needs to extend the existing ROB structure with shadow metadata state to record the information needed to reconstruct the ISA trace.
For example, instead of extracting this information from across modules, we can record the opcode type with the instruction when it enters the ROB and its operand values when it finishes execution.
If the instruction commits later, we can directly obtain the information for the \swcheck.

Nonetheless, even though the shadow logic needs to be customized for different processor designs, the task of deriving shadow logic to extract ISA traces is straightforward for the processor designers to carry out.
We think this presents an important advantage of our scheme compared to related work, as our scheme can get the designers involved in the verification loop and make the formal tools more accessible to architects.
Moreover, our verification scheme is reusable across different design variations, defenses and contracts, as we only need to swap in different ISA trace extraction logic without requiring to search for new invariants as needed by existing work.

Remarkably, there's a scenario where no additional engineering effort is required for tackling different designs. 
Imagine a hardware designer integrating security measures like STT and DoM onto an initially insecure processor.
The shadow logic, initially integrated for the insecure design, seamlessly transfers to the secure design without alteration.
Indeed, all the necessary signals for the shadow logic remain intact within the processor when adding the defense mechanism. 
We use this observation in \cref{sec:eval:defense}, where we assess different defense schemes employing precisely the same shadow logic.




\subsection{Challenges: Two Requirements}
\label{sec:main:twoproperties}



Enforcing the \swcheck needs comparing two ISA-level traces.
However, naively introducing shadow logic to extract committed instruction information is insufficient to verify software-hardware contracts since it will violate what we name the \textbf{instruction inclusion requirement} and the \textbf{synchronization requirement}.

\subsubsection{Instruction Inclusion Requirement}

Recall from \cref{eq:4trace}: the \swcheck and the \hwcheck examine the same program $P$.
This leads to the instruction inclusion requirement: ``\textit{if an instruction's microarchitecture side effects have been considered in the \hwcheck and it will eventually commit, this instruction must also be examined by the \swcheck}.''

The baseline verification scheme assumes the out-of-order processor fetches at most one instruction per cycle and the ISA processor executes one instruction per cycle.
The instruction inclusion requirement is automatically fulfilled as the ISA processor is always running ahead of the out-of-order processor.

However, when switching to our optimized verification scheme, enforcing the instruction inclusion requirement requires additional work.
Indeed, intuitively, in the out-of-order machines, the \swcheck is checked at commit time, because the \swcheck only examines committed instructions. 
This is a bit behind the \hwcheck, meaning we may miss the instructions that are still in their execution or even the fetch stage. 
Thus, we need to make sure that any violation of the \hwcheck is not related to a failure of \swcheck that has not occurred yet, \cameraready{i.e., the violation of the \hwcheck is not for a program that should have been filtered out by the \swcheck}.

\paragraph{Solution}
Our solution to enforce the instruction inclusion requirement is to structure the verification logic into two phases.
In the first phase, we perform the \swcheck and \hwcheck on a pair of out-of-order processors cycle by cycle.
Upon detecting microarchitecture trace deviation, we switch to the second phase, where the \swcheck is applied to all the inflight and bound-to-commit instructions in the pipeline, satisfying the instruction inclusive requirement.

\subsubsection{Synchronization Requirement}

In the baseline verification scheme, for each pair of machines, the two machines in the pair execute in a lock-step manner. 
Hence, the scheme can 
perform both the \swcheck and the \hwcheck by comparing the corresponding states in each pair at \emph{each cycle}.
Because the ISA machine executes exactly one instruction per cycle, the check on ISA traces ($O_{ISA}$) is equivalent to comparing states instruction after instruction.

However, comparing at each cycle the new elements of the ISA traces derived from the two out-of-order processors will not be an equivalent check.
This is because the out-of-order machines might run out of synchronization when the two processors commit their instructions at different time.
Consider the case when the two processors try to commit a load instruction with different addresses, which can happen because of different values of the secret.
One of the processors commits the load in one cycle as the load hits the L1 cache, while the other processor suffers a cache miss and commits the load after the result returns from DRAM, potentially after a hundred cycles.
Because the instruction commits at different cycles, the ISA trace derived from the first processor is no longer cycle-aligned with the ISA trace derived from the second processor.

\paragraph{Solution}
To perform the same check for the \swcheck in the baseline scheme, we first need to \textit{realign} the derived ISA traces.
We achieve this by introducing shadow logic to the second phase of our verification scheme to synchronize the out-of-order processors to commit instructions at the same speed.

\begin{lstlisting}[style=mystyle, language=mylang, basicstyle=\ttfamily\linespread{0.85}\small, label={fig:contract_code}, caption={Verilog-Style Pseudo-code for Shadow Logic. },captionpos=b] 
cpu cpu1 (.clk(pause1 ? 0: clk), .rst(rst)); 
cpu cpu2 (.clk(pause2 ? 0: clk), .rst(rst));

always @(posedge clk) begin
  // shadow logic to compare traces
  uarch_diff = (O_uarch(cpu1)!=O_uarch(cpu2));
  isa_diff   = (O_ISA(cpu1)  !=O_ISA(cpu2));
  
  // phase-1
  initial uarch_diff_phase1 <= 0;
  if (uarch_diff_phase1==0 && uarch_diff) then
    uarch_diff_phase1 <= 1;
    tail1 <= cpu1.tail;
    tail2 <= cpu2.tail;
  endif
 
  // phase-2
  if uarch_diff_phase1 then
    // Requirement 1: inst inclusion
    drained = (tail1==cpu1.head-1) &&
              (tail2==cpu2.head-1);
    
    // Requirement 2: realign ISA traces
    if (cpu1.commit==cpu2.commit)
        pause1 <= 0;
        pause2 <= 0;
    else if (cpu1.commit)
        pause1 <= 1;
    else if (cpu2.commit)
        pause2 <= 1;
  endif

  // contract assumption check
  assume (isa_diff==0); 
  // leakage assertion check
  assert (!(uarch_diff_phase1 && drained)); 
end
\end{lstlisting} 


\subsection{Solution: Two-Phase Shadow Logic}
\label{sec:main:overview}


\cref{fig:contract_code} shows the pseudocode of our two-phase shadow logic written in Verilog style.
Given an out-of-order processor to be verified, we create two instances of the processor, named \texttt{cpu1} and \texttt{cpu2} (lines 1-2).
We then introduce shadow logic to drive and monitor the microarchitecture traces and ISA traces of the two instances cycle by cycle (lines 5-7).

The remaining shadow logic uses the trace comparison results and operates differently in two phases.
In Phase 1, upon a microarchitecture trace deviation, we record the deviation by setting the variable \texttt{uarch\_diff\_phase1} to true and also record the tail of the ROB of each cpu instance.
The goal of recording the tail is to assist the shadow logic in Phase 2 to determine when the \swcheck is completed to satisfy the instruction inclusion requirement.

In Phase 2, since a microarchitecture trace deviation has already happened, the responsibility of the remaining shadow logic is to ensure the \swcheck is sound, satisfying both the instruction inclusion requirement and the synchronization requirement.
Specifically, the shadow logic tracks whether the two instances have committed the instruction that was recorded at the end of Phase 1 (lines 19-21).
Note that the actual shadow logic implementation also accounts for the case when the recorded tail is squashed due to mis-speculation, which we do not show in \cref{fig:contract_code} for brevity.
Besides, the shadow logic also re-aligns the ISA observation traces whenever the two instances go out of synchronization.
The re-alignment operation is achieved by manipulating the input \texttt{clk} signal to the two cpu instances, involving no code modifications inside the cpu code (see lines 23-30 and how \texttt{pause} is used in lines 1-2).

Note that the re-alignment operation only happens in Phase 2 because the commit signal is included in the $O_{\mu arch}$ trace, which is a common practice in contracts targeting timing side channels.
\cameraready{This example shows a machine (the cpu instance) with one clock domain.
In the case that the machine has multiple clock domains, e.g., using different clocks for the processor and caches, all the clocks need to be paused and resumed simultaneously.}

Finally, we use SVA to explicitly express the \swcheck and \hwcheck in lines 33-36.

\paragraph{Model Checking with \name}
Given the cpu state machines and the shadow logic, model-checking techniques check for all possible input states that satisfy the assumption ($M_{pub}$ and $M_{sec}$) and answer the question whether any of the input states can trigger the assertions.
We expect three possible outcomes:
\begin{itemize}[leftmargin=*]
    \item[1)] the model checker finds a counterexample, which reassembles an attack code that satisfies the software constraints but will introduce distinguishable microarchitecture traces running on the cpu;
    \item[2)] the model checker declares a successful unbounded proof to indicate the processor under verification does satisfy the software-hardware contract for unbounded number of cycles;
    \item[3)] the model check hits a scalability bottleneck and cannot produce a result within a reasonable amount of time, which we refer to as ``time out.''
\end{itemize}

\paragraph{\rev{Validity} of the Synchronization Operation}
As discussed in \cref{sec:background:shadow}, shadow logic should only passively monitor operations and never modify the design under verification.
However, the shadow logic we introduce does have a limited influence on the design when the ``\texttt{pause}'' signal freezes the clock. 
Strictly speaking, this logic is no longer just monitoring the design.
\rev{
We claim our shadow logic is still valid for two reasons.
First, the shadow logic performs the ``\texttt{pause}'' operation in Phase 2.
At the beginning of Phase 2, the timing-dependent microarchitectural traces deviation has already been detected, thus Phase 2 only focuses on comparing traces of committed instructions, i.e. ISA traces, which only involve the execution results of committed instructions and does not include any microarchitectural timing information.
Second, the ``\texttt{pause}'' signal only freezes the clock signal of the processor and thus does not affect the execution results of committed instructions.
Thus, it will not influence the ISA traces, nor will it affect the ISA trace comparison results.
}

\paragraph{Supporting Superscalar Processors}
The pseudocode in \cref{fig:contract_code} assumes the processor commits maximally one instruction at a cycle.
Our scheme can be adapted for superscalar processors, where multiple instructions can be committed at each cycle.
The problem that we need to tackle is that the pausing granularity is determined by how many instructions the processor commits.
We will need to add shadow logic to break the atomicity of \swcheck by supporting partial ISA trace comparison at each cycle and the capability to remember unaligned traces for future comparison.
This approach introduces minimal shadow state overhead as the number of entries for memorizing unmatched ISA traces only needs to match the commit bandwidth of the processor. 
It is constant regardless of the ISA trace length.
However, the complexity of the logic increases with the commit bandwidth because we need to align traces across neighboring cycles.

\cameraready{
\subsection{Discussion}

\paragraph{Verification Gap.}
Our verification scheme reformulates the contract property in Equation~\ref{eq:4trace} and can potentially introduce a verification gap.
Our proof is conditioned upon the ISA traces stated as the hypothesis in Equation~\ref{eq:4trace} matching the ISA traces extracted from the out-of-order processors, which in turn requires the following assumptions to be satisfied:
1) the shadow logic is assumed to be implemented correctly;
2) the out-of-order processor is assumed to be functionally correct.

First, the definition of the correctness of our shadow logic includes the shadow logic selecting the right signals to extract the ISA trace and the microarchitectural trace, and the shadow logic generating the pause signal at the proper time.
As such, we ensure the derived ISA trace from the out-of-order processor contains the writeback data of each corresponding instruction.

Second, the functional correctness assumption ensures that the commit stage generates the correct ISA instructions and writeback data as the ISA specification.
Together with the shadow logic correctness assumption, we ensure the derived ISA traces in our scheme are equivalent to the ISA hypothesis in the contract property.

}
\cameraready{
\paragraph{Decoupling of security and functional verification.}
Our verification approach leverages the decoupling (i.e., separation) of functional verification and security verification.
Specifically, our approach for verifying a security property generates a verification task that does not embody a proof obligation related to the functional correctness of the out-of-order processor.
Instead, as stated above, we assume the machine has been independently verified to be functionally correct.
Our choice is justified because verification engineers spend significant effort on functional verification, and our approach leverages this and avoids having to redo this effort as a sub-task of security verification, which may explain the significant verification performance improvement obtained with our approach.
Similar decoupling approaches are also used in inductive approaches on verifying contract properties such as LEAVE~\cite{wang2023specification}.
}


\section{Experiment Setup}
\label{sec:experiment}

\begin{table}[t]
  \setlength\tabcolsep{4pt}
  \small
  \centering
  \begin{tabular}{|c|L{6.7cm}|} \hline
  \multirow{6}{*}{\begin{tabular}{@{}c@{}}\texttt{Sodor}\\\cite{website:sodor}\end{tabular}}
    &Open-source in-order processor \\ 
    &ISA: the full suite of RV32I \\
    &Configuration: 2-stage pipeline, 1-cycle memory \\
    & Code size: 700 lines of Chisel (2,700 lines of Verilog)\\
    & \qquad\qquad\space $\sim$90 lines of Verilog for shadow logic\\
    & \qquad\qquad\space $\sim$3h manual effort for shadow logic\\
    
  \hline
  
  \multirow{7}{*}{\begin{tabular}{@{}c@{}}\texttt{Simple} \\\texttt{OoO}\end{tabular}}
    &Our in-house minimal out-of-order processor \\
    &ISA: 4 customized insts (loadimm, ALU, load, branch) \\
    &Configuration: 4-stage pipeline,  4-entry ROB  \\
    &\qquad\qquad\qquad\space commit bandwidth is 1 inst/cycle \\
    & Code size: 1,000 lines of Verilog\\
    & \qquad\qquad\space $\sim$100 lines of Verilog for shadow logic\\
    & \qquad\qquad\space $\sim$5h manual effort for shadow logic\\
  \hline
  
  \multirow{7}{*}{\begin{tabular}{@{}c@{}}\texttt{Ride}\\\texttt{core}\\\cite{website:ridecore}\end{tabular}} 
    &Open-source out-of-order processor \\
    &ISA: 35 instructions in RV32IM \\
    &Configuration: 6-stage pipeline, 8-entry ROB \\
    &\qquad\qquad\qquad\space commit bandwidth is 2 inst/cycle\\
    & Code size: 8,100 lines of Verilog\\
    & \qquad\qquad\space $\sim$400 lines of Verilog for shadow logic\\
    & \qquad\qquad\space $\sim$10h manual effort for shadow logic\\
  \hline
  \multirow{7}{*}{\begin{tabular}{@{}c@{}}\texttt{BOOM}\\\cite{zhaosonicboom}\end{tabular}} 
    &Open-source out-of-order processor \\
    &ISA: RV64GC \\
    &Configuration: SmallBoomConfig \\
    &\qquad\qquad\qquad\space 32-entry ROB \\
    &\qquad\qquad\qquad\space commit bandwidth is 1 inst/cycle \\
    & Code size: 19k lines of Chisel (136k lines of Verilog)\\
    & \qquad\qquad\space $\sim$240 lines of Verilog for shadow logic\\
    & \qquad\qquad\space $\sim$8h manual effort for shadow logic\\
    \hline
  \end{tabular}
  \caption{Detailed processor configurations.}
  \label{tab:eval:setup}
\end{table}


We implement shadow logic for verifying the sandboxing contract and the constant-time contract (\cref{sec:background:contract}) on four processors as shown in \cref{tab:eval:setup}.
Other than the three open-source processors, SimpleOoO is a toy out-of-order processor developed by us that has the basic out-of-order execution behaviors, and we augment it with different defense mechanisms.
In \cref{tab:eval:setup}, we compare the code size for each processor that we verified and the code size and manual effort for the shadow logic.
Note that although BOOM is much larger than Ridecore, its shadow logic is less complex because it only commits one instruction every cycle while Ridecore is a superscalar processor that can commits two instructions per cycle.
This shows that the complexity of our shadow logic does not necessarily grow as the processor gets larger.


Here is the overview of our verification workflow:
\begin{enumerate}[leftmargin=*]
    \item[1.] Implement the \name that extracts and compares the information from two out-of-order processors.
    The design under verification (DUV) includes both processors and the shadow logic.
    \item[2.] Initialize the instruction memories in two out-of-order processors to have symbolic values, so all possible programs will be explored in model checking.
    \item[3.] Write assumptions (constraints) including: (1) the two out-of-order processors have the same initial state except that the secret in memory has different initial values; (2) the \swcheck must always hold.
    \item[4.] Model check the DUV with the above assumptions and the assertion that \hwcheck holds.
\end{enumerate}

We perform verification tasks using the commercial model checking tool, JasperGold~\cite{website:jaspergold}.
We use its \texttt{Mp} and \texttt{AM} solving engines to find proofs, and use the \texttt{Ht} engine to detect leakage.
The performance results were obtained on a server with an Intel Xeon E5-4610 processor running at 2.4 GHz.
We set 7 days as the time-out period for each verification task.






\section{Experiment Results}
\label{sec:eval}



\subsection{Comparison with Baseline and Existing Schemes}
\label{sec:eval:compare}

\begin{table}[t]
  \setlength\tabcolsep{2pt}
  \centering
  \small

  \begin{tabular}{|l|c|c|c|c|c|} \hline
    & \multicolumn{5}{c|}{\textbf{Sandboxing Contract}} \\ \cline{2-6}
                  
    & \multicolumn{1}{c|}{\texttt{Sodor}}   & \multicolumn{1}{c|}{\texttt{SimpleOoO-S}}   & \multicolumn{1}{c|}{\texttt{SimpleOoO}}   & \multicolumn{1}{c|}{\texttt{Ridecore}}   & \multicolumn{1}{c|}{\texttt{BOOM}} \\ \hline\hline
                  
    Baseline
    &\timeout &\timeout &\bug &\bug &\graycell \\ \hline
  
    LEAVE~\cite{wang2023specification}
    &\secure &\falsealert &\falsealert &\graycell &\graycell  \\ \hline
  
    UPEC~\cite{fadiheh2020formal}
    &\graycell &\graycell &\graycell &\graycell &(\bug) \\ \hline
  
    Our Scheme
    &\secure &\secure &\bug &\bug &\bug \\ \hline
  \end{tabular}

  \caption{A summary of comparing the Baseline, LEAVE~\cite{wang2023specification}, UPEC~\cite{fadiheh2020formal}, and our scheme on verifying the sandboxing contract on 5 processors. ``\bug'' indicates valid attacks are found (``(\bug)'' indicates UPEC cannot find all possible attacks on \texttt{BOOM}), ``\secure'' indicates a proof is found, ``\timeout'' indicates a time out on the verification task {after running for 7 days}, ``\falsealert'' indicates false counterexamples are found, and a \colorbox{lightgray}{shaded} entry indicates the experiment is not conducted.}
  
  \label{tab:eval:all}
\end{table}

                  
                  
  
  
  


To better understand how our work compares to the existing techniques, we conduct experiments to compare our scheme with the baseline, LEAVE~\cite{wang2023specification}, and UPEC~\cite{fadiheh2020formal}.
\cameraready{
We also include the baseline scheme, which is directly derived from \cref{eq:4trace}.
}
We compare the verification results on five processor designs, two secure ones (\texttt{Sodor} and \texttt{SimpleOoO-S}) and three insecure ones (\texttt{SimpleOoO}, \texttt{Ridecore}, and \texttt{BOOM}), regarding two contracts, the sandboxing contract and the constant-time programing contract.
The secure out-of-order core, noted as \texttt{SimpleOoO-S}, delays the issue time of a memory instruction until its commit time if at the time when it enters the pipeline, there is a branch before it in the ROB.

\subsubsection{Comparison Summary}
The comparison results regarding the sandboxing contract are summarized in \cref{tab:eval:all}.
We observe similar results regarding the constant-time contract except UPEC, which does not support it.
Across all five processors, our scheme can successfully find attacks or proofs within a reasonable amount of time (concrete numbers reported later).

\begin{itemize}[leftmargin=*]
    \item Compared to the baseline, our scheme has a clear advantage in conducting unbounded proofs, where the baseline easily times out on finding proofs on secure designs.
    We observe similar performance in finding attacks between the baseline and our scheme.
    \item Similar to our scheme, LEAVE can find proof on in-order processors (\texttt{Sodor}). 
    More importantly, our scheme performs better than LEAVE on out-of-order processors.
    While LEAVE struggles with \texttt{SimpleOOO} designs, generating false counterexamples for both secure and insecure designs, our approach can find bugs and prove security of these designs.
    \item Since UPEC is a highly customized verification scheme and requires advanced formal-method expertise to be adapted to a different design, we only show evaluation results from their paper. Even though it can successfully find attacks on \texttt{BOOM}, the approach falls short in finding all possible attacks, since it can only find attacks whose speculation source is explicitly specified by the user.
    \item \cameraready{Our scheme still cannot scale to fully proving the security of large out-of-order processors such as patched Ridecore and BOOM, nor do other schemes (UPEC can do a partial security proof for BOOM due to their customization).
    }
\end{itemize}


We provide a detailed comparison below.

\subsubsection{Comparison to Baseline}
When verifying a secure design (the first two columns), the baseline scheme times out, unable to generate an answer.
In contrast, our scheme can successfully find the proof within 151 minutes.
\cameraready{
In formal verification, a timeout indicates that the model checker is hitting an exponential wall on the model being verified, i.e., the exponential size of the reachable state space is beyond the reach of the solver.
}
This result highlights that our scheme shows much better scalability of RTL verification for software-hardware contracts compared with the baseline scheme.

\subsubsection{Comparison with LEAVE}
LEAVE uses an inductive approach to verify processors against software-hardware contracts.
Their key contribution is to automatically derive inductive invariants to assist deriving an inductive proof.
These invariants are supposed to capture the reachable states of the state machines being verified.
They start with a set of automatically generated or manually written candidate invariants and use a search algorithm to iteratively remove invariants that do not hold.
The expectation is that the final remaining invariants are strong enough to prove security.
Otherwise, {when insufficient invariants are used, the induction procedure will consider \textit{unreachable} states and generate false counterexamples}.
As such, they declare the verification result as \textit{UNKNOWN}.
An UNKNOWN result indicates that either the processor may be secure, but the search algorithm cannot find a good set of invariants to make the proof go through, or the processor may be insecure. 
{As a side note, LEAVE proposes invariants that consider the {instruction inclusion} and {synchronization} requirements (\cref{sec:main:twoproperties}) in a limited way for in-order processors. 
Our scheme is the first to satisfy these two requirements in a general way, supporting out-of-order and superscalar processors.}

LEAVE has only been used to verify in-order processors.
As LEAVE is open-sourced~\cite{wang2023specification-code}, we extend the evaluation of LEAVE to out-of-order processors.
We use the automatically generated candidate invariants described in LEAVE's paper, i.e., values in corresponding registers are equivalent in the two copies of out-of-order processors.

When verifying an insecure version of \texttt{SimpleOoO}, our scheme finds an attack in 2 seconds.
However, we observe LEAVE fails to find a set of invariants to prove security and states that the result is UNKNOWN, i.e., LEAVE cannot tell whether the design is secure or not.
When verifying \texttt{SimpleOOO-S} (the secure version), our scheme is able to complete the proof within 2.5 hours, while LEAVE reports UNKNOWN after searching invariants for 2.1 hours.
During its execution, we observe that LEAVE's invariant-search algorithm continuously removes invariants from the candidate sets until there are none left.
The reason is that the automatically generated candidate invariants are insufficient to cover reachable states for even simple out-of-order processors.
In these cases, LEAVE will require additional effort to improve the invariant generation algorithm or ask the user to manually write invariants, both posing open research challenges.


\subsubsection{Comparison with UPEC}
UPEC uses an inductive verification approach, assisted with substantial manual invariants {by formal-method experts}, to verify complex out-of-order processors including \texttt{BOOM}.
As discussed in \cref{sec:intro}, UPEC's unbounded proof of secure designs is closely coupled with the defense scheme to be verified. 
{Specifically, they verified a defense scheme that prevents speculative load instructions to execute before all preceding branch instructions in the ROB are resolved.
With such a defense, under the sandboxing contract, the secret data never propagates to any other gates in the design.
Their induction procedure finds that this invariant holds for arbitrary initial states and they can complete an unbounded proof by proving an induction step that only involves 1 cycle of symbolic execution.
}
However, this invariant does not hold for most defenses proposed in the computer architecture community where speculative states do spread in a complex manner inside the processor, as with DoM~\cite{sakalis2019efficient}, STT~\cite{yu2019speculative}, and others.

In addition, we find their approach achieves high scalability at the cost of losing generality and completeness.
Specifically, UPEC's verification procedure requires the user to provide information on \textit{the source of speculation} in the processor.
Their open-source prototype uses branch misprediction as the sole source of speculation, and their manual invariants were developed based on this assumption.


To compare our scheme with UPEC, we implemented our \name on \texttt{BOOM} for both sandboxing and constant-time contracts (UPEC only handles the sandboxing contract).
We use the \texttt{SmallBOOM} configuration and directly use the data cache and instruction cache as the memory with the virtual memory module turned off.
Both cache sizes are configured to 256 bytes.

Our experiment results show that our scheme can find multiple attacks on \texttt{BOOM} exploiting different speculation sources.
First, we find an attack for the sandboxing contract in 120 minutes and a similar attack for the constant-time contract in 36.6 minutes.
In these two attacks, mis-speculation is triggered by exceptions due to memory address misalignment:
\begin{minipage}{\columnwidth}
\begin{center}
\begin{tabular}{c}
\begin{lstlisting}[style=mystyle, language=mylang, label={lst:future}]
lui r6, 65536 //set addr to a legal range
lhu r14, 1221(r6) //mis-alignment in addr
lh r0, r14    //use secret as load addr
\end{lstlisting}
\end{tabular}
\end{center}
\end{minipage}
In the above code snippet, the second instruction loads the secret into \texttt{r14} and the third instruction uses the secret as memory address, thus violating the \hwcheck.
Since the second instruction triggers an exception due to address misalignment, the second and third instructions will not be committed, and thus will satisfy the \swcheck. 
Note that, these attacks were not reported by UPEC on \texttt{BOOM} as they only consider branch misprediction as the speculation source. 


Second, we can continue to search for other attacks following the standard practice in formal verification.
We add an assumption to exclude the first attack that we found.
The added assumption ensures the input program does not involve memory accesses using misaligned addresses.
Running the verification procedure for the sandboxing contract leads us to the discovery of a new attack exploiting exceptions due to illegal memory accesses.
The verification for the constant-time contract generates an attack exploiting branch misprediction.
The former case takes 8.7 hours, and the latter case takes 1.4 hours.

Finally, we further continue the experiments to exclude the two attacks.
The verification tasks time out after 24 hours.

The above experiments show that our scheme can scale to find attacks on complex out-of-order cores, such as \texttt{BOOM}.
In addition, 
our scheme finds unknown attacks on processors without specifying the source of speculation and ends up finding multiple attacks that UPEC was unable to find.

\subsection{Impact of Defense Mechanisms}
\label{sec:eval:defense}

\begin{table}[t]
  \setlength\tabcolsep{2pt}
  \small
  \centering
  
                  
                  
  
  
  
  

  
  \begin{tabular}{|l|R{2.2cm}|R{2.2cm}|} \hline
                  & \multicolumn{1}{c|}{\textbf{Sandboxing}} & \multicolumn{1}{c|}{\textbf{Constant-Time}}  \\ \hline\hline
                  
  $\mathsf{NoFwd_{futuristic}}$ &\textcolor{proof}{66min}  &\textcolor{bug}{0.4s}     \\ \hline
  
  $\mathsf{NoFwd_{spectre}}$    &\textcolor{proof}{45h}    &\textcolor{bug}{0.1s}     \\ \hline
  
  $\mathsf{Delay_{futuristic}}$ &\textcolor{proof}{21min}  &\textcolor{proof}{10min}  \\ \hline
  
  $\mathsf{Delay_{spectre}}$    &\textcolor{proof}{151min} &\textcolor{proof}{37min}  \\ \hline
  
  $\mathsf{DoM_{spectre}}$      &\textcolor{bug}{6.5min{\color{black}\footnotemark[2]}}
                                                           &\textcolor{bug}{5.9min}   \\ \hline

  \end{tabular}

  \caption{Verification time using our scheme on \texttt{SimpleOoO} augmented with different defenses. The \textcolor{bug}{red entry} indicates an attack is found and the \textcolor{proof}{green entry} indicates a proof is found.}
  \label{tab:eval:compare-defense}
\end{table}

\footnotetext[2]{These two attacks are found using an 8-entry ROB instead of the default 4-entry ROB because these attacks require more concurrent instructions.}

{We now study how various factors can affect the verification time of our scheme.
We start with the influence of defense mechanisms on verification time.}
We extend the \texttt{SimpleOoO} core with five microarchitecture defenses from the literature~\cite{sakalis2019efficient,yu2019speculative,weisse2019nda}.
Note that we can directly reuse the shadow logic we developed for \texttt{SimpleOoO}, highlighting the reusability of our approach.
We note this is the \textit{first} time for many of these defense proposals to be evaluated as RTL designs, in contrast to high-level abstract models~\cite{guarnieri2021hardware, yang2023pensieve, trippel2018checkmate} or via gem5 simulation~\cite{gem5}.

\paragraph{Defenses}
The following two mechanisms share similarities with the defenses described in STT~\cite{yu2019speculative} and NDA~\cite{weisse2019nda}.
\begin{itemize}[leftmargin=*]
  \item $\mathsf{NoFwd_{futuristic}}$: Do not forward the data of a load instruction to younger instructions until its commit time.
  \item $\mathsf{NoFwd_{spectre}}$: Do not forward the data of a load instruction to younger instructions until its commit time, if 
  when the load enters the pipeline, there is a branch before the load in the ROB.
\end{itemize}
The difference between the two mechanisms is that $\mathsf{NoFwd_{futuristic}}$ considers all possible sources of speculation, while $\mathsf{NoFwd_{spectre}}$ only considers branch prediction as the source of speculation.

The next three defenses work by delaying load instructions at different points in their lifetime.
\begin{itemize}[leftmargin=*]
  \item $\mathsf{Delay_{futuristic}}$: Delay the issue time of a memory instruction until its commit time.
  \item $\mathsf{Delay_{spectre}}$: Delay the issue time of a memory instruction until its commit time, if at the time when it enters the pipeline, there is a branch before it in the ROB.
  It is noted as \texttt{SimpleOoO-S} in \cref{sec:eval:compare}.
  \item $\mathsf{DoM_{spectre}}$: This models a simplified version of the Delay-on-Miss defense~\cite{sakalis2019efficient}.
  It always speculatively issues a load instruction. If the load hits the L1 cache, return the data to the core and proceed.
  If the load misses the L1 cache, delay the load from being sent to the main memory if at the time when the load enters the pipeline, there is a branch before the load in the ROB.
  We model a cache with a single cache entry with a 1-cycle hit and a 3-cycle miss.
\end{itemize}
$\mathsf{Delay_{futuristic}}$ and $\mathsf{Delay_{spectre}}$ are secure on \texttt{SimpleOoO} for both the sandboxing contract and constant-time contract, while $\mathsf{DoM_{spectre}}$ is not secure.
The vulnerability of $\mathsf{DoM_{spectre}}$ has been discovered in prior work~\cite{behnia2021speculative,fustos2020spectrerewind}.

\paragraph{Evaluation Results}
\cref{tab:eval:compare-defense} compares the checking time of verifying the five defenses.
For all the cases where we are supposed to find a proof, i.e., when the designs are secure, our scheme can find a proof \rev{for all of them within 45 hours, with most completing in about an hour.}
We note in all the cases, the baseline scheme times out (not shown in the table). 

Comparing the checking time for different defenses, we observe that defenses have a non-trivial impact on checking time.
First, it takes much longer to find the proof in secure designs (above 10 minutes) than to find an attack in insecure designs (below 10 minutes).
Second, it takes less time to find a proof on a relatively more conservative defense mechanism.
Comparing the pair of \texttt{NoFwd} and the pair of \texttt{Delay} defenses, the futuristic version takes much less time to verify.

\subsection{Impact of Structure Sizes}
\label{sec:eval:size}

\begin{figure}[t]
    \centering
    \includegraphics[width=\columnwidth]{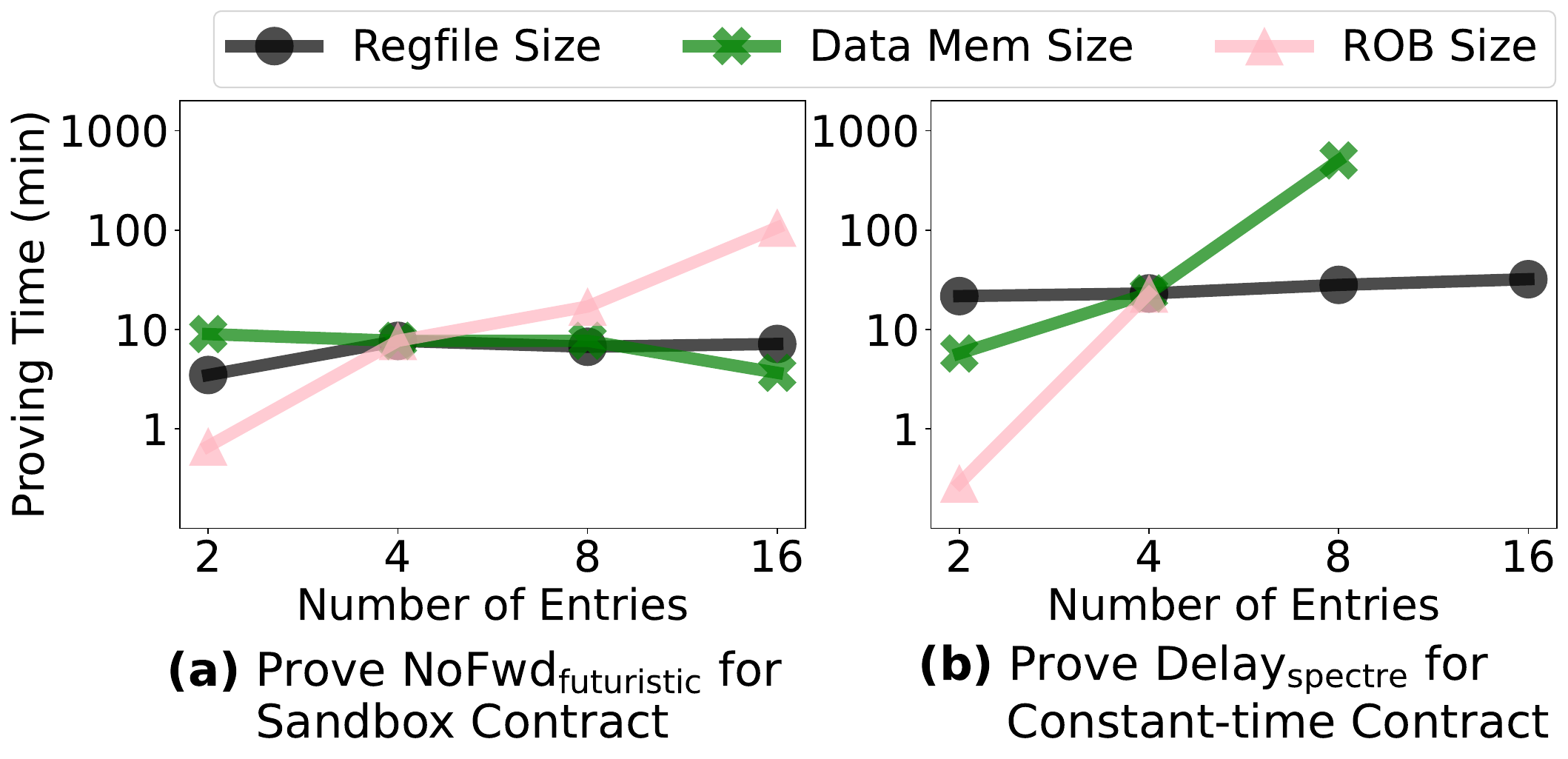}
    \caption{Verification time when varying the size of the register file, data memory, and re-order buffer. }
    \label{fig:eval:size-time}
\end{figure}

Finally, we study how structure sizes of re-order buffer, register file, and memory affect  verification time.
\cref{fig:eval:size-time} shows the checking time for verifying $\mathsf{NoFwd_{futuristic}}$ defense for the sandboxing contract and $\mathsf{Delay_{spectre}}$ defense for the constant-time contract.
The default configuration consistently uses 4 entries for the register file, ROB, and data memory.
The three lines in the plot are derived by changing one of the three structures and keeping the other two unchanged.

Overall, the size of the register file has negligible impact on verification time.
The size of data memory has a limited impact on proving the sandboxing contract but has a larger impact on proving the constant-time contract.
Note that the size of ROB has a significant impact (y-axis is log scale).
When we increase the ROB size, the verification time increases exponentially.
The ROB size determines the number of inflight instructions, and the space of possible interactions between instructions grows quickly with ROB size, taking a much longer time to check.

\section{Limitations and Future Work}
\label{sec:discussions}

Despite being able to achieve significant verification performance improvement compared to the baseline and several related works, we acknowledge this paper has several limitations.
First, one appealing feature of our verification scheme is that the proposed contract shadow logic is \textit{conceptually} reusable across different designs, given that the primary task of the shadow logic is to simply extract committed instruction information.
We must concede that, thus far, constructing and synchronizing the shadow logic for the designs in our evaluation has relied on manual effort.
To further reduce the required manual effort, we want to investigate how to partially automate this process.
Our problem shares similarity to \emph{tandem verification}~\cite{Xing2022Generalizing} for functional correctness, where the same ISA-level information is extracted from RTL implementations. 
Applying techniques similar to~\cite{Xing2022Generalizing} might help in automating part of the effort.

Second, our scheme does not fully address the scalability issue for verifying out-of-order processors with a realistic ROB size, which is usually above 100.
This scalability issue is a well-known problem in hardware verification.
On the positive side, we believe a reasonably reduced-size out-of-order machine already exhibits the essential security-relevant features of a real-size out-of-order machine, including speculation, resource contention, and instruction reordering.
However, we believe work can be done to further improve the scalability. 
One promising direction is to explore taint propagation techniques like GLIFT~\cite{tiwari2009complete,solt2022cellift}.
As studied in prior work~\cite{hu2011theoretical}, taint propagation requires making a delicate trade-off between precision and verification overhead. 
The taint propagation logic can be customized to the security property and specific hardware design to accomplish this.
We hypothesize that taint propagation techniques could be combined with our scheme to further improve verification performance.

Finally, as we explore the relationship between manually crafted invariants and and the information embedded within our shadow logic, some additional insights emerge.
We speculate that our shadow logic may offer a rich array of invariants to the model checker.
This hypothesis is rooted in the notion that contract constraints, originally designed to constrain software programs, indirectly strongly constrain the set of reachable states of the processor.
However, more investigation is needed to understand the constraining power of the shadow logic and how this influences the performance of the verification procedure.
\section{Related Work}
\label{sec:related}


Among all related works, UPEC~\cite{fadiheh2020formal} and LEAVE~\cite{wang2023specification} are the closest to us, as they both target speculative execution vulnerabilities and work on RTL.
We have provided detailed comparisons of our scheme, UPEC, and LEAVE in \cref{sec:eval:compare}.
We now discuss other related work.
We identify that previous approaches differ along three axes: 1) verification target, 2) security properties, and 3) checking techniques.

\paragraph{Targets for Verification of Secure Speculation}
There is a lot of work leveraging formal methods to verify secure speculation properties (variants of software-hardware contracts), but their targets differ from our work, either focusing on abstract processor models or software.

Pensieve~\cite{yang2023pensieve}, Guarnieri et al.~\cite{guarnieri2021hardware}, and CheckMate~\cite{trippel2018checkmate} target abstract processor models (not RTL).
Specifically, Pensieve~\cite{yang2023pensieve} does bounded model checking to detect speculative execution vulnerabilities on modular models connected using hand-shaking interfaces.
Guarnieri et al.~\cite{guarnieri2021hardware} uses a paper proof to reason about an abstract out-of-order processor model expressed as operational semantics.
CheckMate~\cite{trippel2018checkmate} applies security litmus tests on abstract processor models expressed as $\mu$hb graphs to synthesize exploit patterns, so as to find new attacks through these patterns. 
Although verifying abstract processor models can give designers key insights, this approach unavoidably neglects RTL implementation details, which may cause speculative execution attacks.

On the software side, there are efforts to guarantee that a program does not contain gadgets that can be exploited by different speculative execution attacks~\cite{cauligi2020constant,cheang2019formal,mosier2022axiomatic,wang2019oo7,wu2019abstract,mosier2023serberus}.
For example, Serberus~\cite{mosier2023serberus} detects code patterns, at compile time, in cryptographic primitives that can be exploited by broad ranges of speculative execution attacks.
This line of work usually assumes a specific hardware model, and then verifies that specific pieces of software satisfy the software constraint component in software-hardware contracts.
The goal is very different from our work, where we verify the hardware's RTL implementation.

\paragraph{Security Properties}
There exist RTL security verification works that target security properties \textit{not} related to speculative execution.
These works target various data-oblivious properties.
For example, Clepsydra~\cite{ardeshiricham2017clepsydra} verifies hardware modules, such as ALUs and Caches, take the same amount of time when executing on different input data.
IODINE~\cite{gleissenthall2019iodine}, Xenon~\cite{v2021solver}, UPEC-DIT~\cite{deutschmann2023scalable} check whether each instruction's execution time is independent of its operands for in-order/out-of-order processors.
Knox~\cite{athalye2022verifying} builds a framework to verify that when a specific program is running on a physically-isolated Hardware Security Module (HSM), its execution time does not leak extra information beyond the functional specification of the program.

\paragraph{Checking Techniques}
In addition to leveraging formal-method techniques, fuzzing-based schemes~\cite{ghaniyoun2021introspectre,moghimi2020medusa,xiao2019speechminer,weber2021osiris,oleksenko2022revizor,oleksenko2023hide,hur2022specdoctor}, such as SpecDoctor~\cite{hur2022specdoctor}, check RTL or blackbox hardware for speculative execution vulnerabilities by generating and testing random attack patterns.
Although the above works cannot guarantee full coverage of speculative execution attacks, they can be faster at detecting security bugs when evaluating large designs.


\paragraph{Verification of Clean-slate Designs}
Finally, an exciting yet orthogonal line of work focuses on security verification of clean-slate designs.
SecVerilog~\cite{zhang2015hardware} enables hardware programmers to enforce security policies while developing the hardware.
A violation of the security policy would be detected at compile time.
SpecVerilog~\cite{zagieboylo2023specverilog} further extends this ability to support specifying speculation-related policies.
\cite{tan2023security} formally verifies that a low-trust architecture with a specialized ISA and hardware is free of timing side-channel leakage.

\section{Conclusions}
\label{sec:conclusion}
Our paper presents \name, a broadly-applicable approach for performing formal verification of security properties on RTL designs for software-hardware contracts.
Our scheme asks processor designers to construct shadow logic to extract ISA observation traces from processors and enforce the instruction inclusion and synchronization requirements.
Our evaluation shows the strength of our scheme compared with the baseline and two state-of-the-art RTL verification schemes in both finding attacks on insecure designs and deriving unbounded proofs for secure designs.


\appendix

\bibliographystyle{plain}
\bibliography{refs}

\end{document}